# NEW PROXIMITY ESTIMATE FOR INCREMENTAL UPDATE OF NON-UNIFORMLY DISTRIBUTED CLUSTERS

A.M.Sowjanya and M.Shashi

Department of Computer Science and Systems Engineering,
College of Engineering, Andhra University, Visakhapatnam, India

*ABSTRACT*

*The conventional clustering algorithms mine static databases and generate a set of patterns in the form of clusters. Many real life databases keep growing incrementally. For such dynamic databases, the patterns extracted from the original database become obsolete. Thus the conventional clustering algorithms are not suitable for incremental databases due to lack of capability to modify the clustering results in accordance with recent updates. In this paper, the author proposes a new incremental clustering algorithm called CFICA(Cluster Feature-Based Incremental Clustering Approach for numerical data) to handle numerical data and suggests a new proximity metric called Inverse Proximity Estimate (IPE) which considers the proximity of a data point to a cluster representative as well as its proximity to a farthest point in its vicinity. CFICA makes use of the proposed proximity metric to determine the membership of a data point into a cluster.*

## 1. INTRODUCTION

Clustering discovers patterns from a wide variety of domain data, thus many clustering algorithms were developed by researchers. The main problem with the conventional clustering algorithms is that, they mine static databases and generate a set of patterns in the form of clusters. Numerous applications maintain their data in large databases or data warehouses and many real life databases keep growing incrementally. New data may be added periodically either on a daily or weekly basis. For such dynamic databases, the patterns extracted from the original database become obsolete. Conventional clustering algorithms handle this problem by repeating the process of clustering on the entire database whenever a significant set of data items are added. Let $S_D$ be the original data base (static database) and $\Delta S_D$ be the incremental database. Conventional clustering algorithms process the expanded database ($S_D + \Delta S_D$) to form new cluster solution from scratch. The process of re-running the clustering algorithm on the entire

DOI : 10.5121/ijdkp.2013.3509  91

dataset is inefficient and time-consuming. Thus most of the conventional clustering algorithms are not suitable for incremental databases due to lack of capability to modify the clustering results in accordance with recent updates.

In this paper, the author proposes a new incremental clustering algorithm called CFICA (Cluster Feature-Based Incremental Clustering Approach for numerical data) to handle numerical data. It is an incremental approach to partitional clustering. CFICA uses the concept of Cluster Feature (CF) for abstracting out the details of data points maintained in the hard disk. At the same time Cluster Feature provides all essential information required for incremental update of a cluster. Most of the conventional clustering algorithms make use of Euclidean distance ( $E_D$ ) between the cluster representatives ( mean / mode / medoid ) and the data point to estimate the acceptability of the data point into the cluster.

In the context of incremental clustering while adopting the existing patterns or clusters to the enhanced data upon the arrival of a significant chunk of data points, it is often required to elongate the existing cluster boundaries in order to accept new data points if there is no loss of cluster cohesion. The author has observed that the Euclidean distance (ED) between the single point cluster representative and the data point will not suffice for deciding the membership of the data point into the cluster except for uniformly distributed clusters. Instead, the set of farthest points of a cluster can represent the data spread within a cluster and hence has to be considered for formation of natural clusters. The authors suggest a new proximity metric called Inverse Proximity Estimate (IPE) which considers the proximity of a data point to a cluster representative as well as its proximity to a farthest point in its vicinity. CFICA makes use of the proposed proximity metric to determine the membership of a data point into a cluster.

## 2. RELATED WORK

Incremental clustering has attracted the attention of the research community with Hartigan's Leader clustering algorithm [1] which uses a threshold to determine if an instance can be placed in an existing cluster or it should form a new cluster by itself. COBWEB [2] is an unsupervised conceptual clustering algorithm that produces a hierarchy of classes. Its incremental nature allows clustering of new data to be made without having to repeat the clustering already made. It has been successfully used in engineering applications [3]. CLASSIT [4] is an alternative version of COBWEB. It handles continuous or real valued data and organizes them into a hierarchy of concepts. It assumes that the attribute values of the data records belonging to a cluster are normally distributed. As a result, its application is limited. Another such algorithm was developed by Fazil Can to cluster documents [5]. Charikar et al. defined the incremental clustering problem and proposed a incremental clustering model which preserves all the desirable properties of HAC (hierarchical agglomerative clustering) while providing a extension to the dynamic case. [6].

BIRCH (Balanced Iterative Reducing and Clustering using Hierarchies) is especially suitable for large number of data items [7]. Incremental DBSCAN was presented by Ester et al., which is suitable for mining in a data warehousing environment where the databases have frequent updates [8]. The GRIN algorithm, [9] is an incremental hierarchical clustering algorithm for numerical data sets based on gravity theory in physics. Serban and Campan have presented an incremental algorithm known as Core Based Incremental Clustering (CBIC), based on the k-means clustering method which is capable of re-partitioning the object set when the attribute set changes [10]. The new demand points that arrive one at a time have been assigned either to an existing cluster or a

International Journal of Data Mining & Knowledge Management Process (IJDKP) Vol.3, No.5, September 2013

newly created one by the algorithm in the incremental versions of Facility Location and k-median to maintain a good solution [11].

## 3. FUNCTIONALITY OF CFICA

An incremental clustering algorithm has to perform the primary tasks namely, initial cluster formation and their summaries, acceptance of new data items into either existing clusters or new clusters followed by merging of clusters to maintain compaction and cohesion. CFICA also takes care of concept-drift and appropriately refreshes the cluster solution upon significant deviation from the original concept.

It may be observed that once the initial cluster formation is done and summaries are represented as Cluster Features, all the basic tasks of the incremental clustering algorithm CFICA can be performed without requiring to read the actual data points ( probably maintained in hard disk ) constituting the clusters. The data points need to be refreshed only when the cluster solution has to be refreshed due to concept-drift.

### 3.1 Initial clustering of the static database

The proposed algorithm CFICA is capable of clustering incremental databases starting from scratch. However, during the initial stages refreshing the cluster solution happens very often as the size of the initial clusters is very small. Hence for efficiency reasons the author suggests to apply a partitional clustering algorithm to form clusters on the initial collection of data points ($S_D$). The author used the k-means clustering algorithm for initial clustering to obtain $k$ number of clusters as it is the simplest and most commonly used partitional clustering algorithm. Also k-means is relatively scalable and efficient in processing large datasets because the computational complexity is O(nkt) where n is the total number of objects, k is the number of clusters and t represents the number of iterations. Normally, k<<n and t<<n and hence O(n) is taken as its time complexity [12].

### 3.2 Computation of Cluster Feature (CF)

CFICA uses cluster features for accommodating the essential information required for incremental maintenance of clusters. The basic concept of cluster feature has been adopted from BIRCH as it supports incremental and dynamic clustering of incoming objects. As CFICA handles partitional clusters as against hierarchical clusters handled by BIRCH, the original structure of cluster feature went through appropriate modifications to make it suitable for partitional clustering.

The Cluster Feature ($CF$) is computed for every cluster $c_i$ obtained from the k-means algorithm. In CFICA, the Cluster Feature is denoted as,

$$CF_i = \{n_i, \vec{m_i}, \vec{m_i'}, \vec{Q_i}, \vec{ss_i}\}$$

where $n_i \rightarrow$ number of data points,

International Journal of Data Mining & Knowledge Management Process (IJDKP) Vol.3, No.5, September 2013$\vec{m_i}$ → mean vector of the cluster $C_i$ with respect to which farthest points are calculated,

$\vec{m_i'}$ → new mean vector of the cluster $C_i$ that changes due to incremental updates,

$Q_i$ → list of p-farthest points of cluster $C_i$

$\vec{ss_i}$ → squared sum vector that changes during incremental updates.

A Cluster Feature is aimed to provide all essential information of a cluster in the most concise manner. The first two components $n_i$ and $\vec{m_i}$ are essential to represent the cluster prototype in a dynamic environment. $n_i$ will be incremented whenever new data point is added to a cluster. $\vec{m_i'}$, the new mean is essential to keep track of dynamically changing nature / concept – drift occurring in the cluster while it is growing. It is updated upon inclusion of a new data point.

The $Q_i$, set of p-farthest points of cluster $C_i$ from its existing mean $\vec{m_i}$, are used to handle non-uniformly distributed and hence irregularly shaped clusters; The set of p - farthest points of the i$^{th}$ cluster, are calculated as follows: First Euclidean distances are calculated between the data points within cluster $c_i$ and the mean of the corresponding cluster $\vec{m_i}$. Then, the data points are arranged in descending order with the help of the measured Euclidean distances. Subsequently, the top p - farthest points for every cluster are chosen from the sorted list and these points are known as the p - farthest points of the cluster $c_i$ with respect to the mean value, $\vec{m_i}$. Thus a list of p - farthest points is maintained in the Cluster Feature for every cluster, $C_i$. These p - farthest points are subsequently used for identifying the farthest point, $q_i$. $\vec{ss_i}$, is the squared sum which is essential for estimating the quality of cluster in terms of variance of data points from its mean

In general, the variance of a cluster ($\sigma^2$) containing 'N' data points is defined as

$$\sigma^2 = \frac{1}{N} \sum_{j=1}^{N} (x_j - \bar{x})^2 = (\frac{1}{N} \sum_{j=1}^{N} x_j^2) - \bar{x}^2 \qquad .$$

In the present context, the error associated with i$^{th}$ cluster represented by its Cluster Feature, $CF_i$ is calculated as given

$$\sigma_i^2 = \left(\frac{1}{n_i} * \vec{SS_i}\right) - \vec{m_i}^2$$

### 3.3 Insertion of a new data point

Data points in a cluster can either be uniformly or non-uniformly distributed. The shape of a uniformly distributed cluster is nearly globular and its centroid is located in the middle (geometrical middle). Non-uniformly distributed clusters have their centroid located in the midst of dense area, especially if there is a clear variation in the density of data points among dense and sparse areas of the cluster. The shape of such clusters is not spherical and the farthest points of a non-uniformly distributed cluster are generally located in the sparse areas.

94

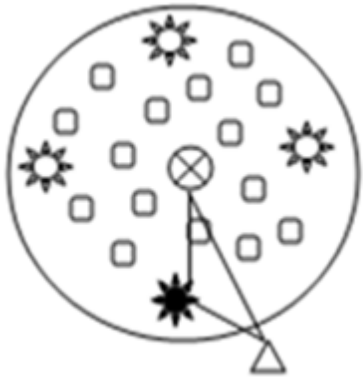 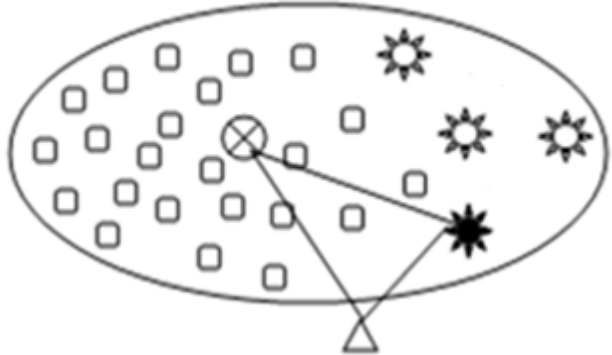

Fig.1 Uniformly Distributed cluster    Fig.2 Non-Uniformly Distributed cluster

Here,   ▫ refers to entities in a cluster

⊗ is the cluster representative ( mean / mode / medoid )

☼ represent p-farthest points of a cluster

✹ farthest point in the vicinity of the entity

△ new entity to be incorporated

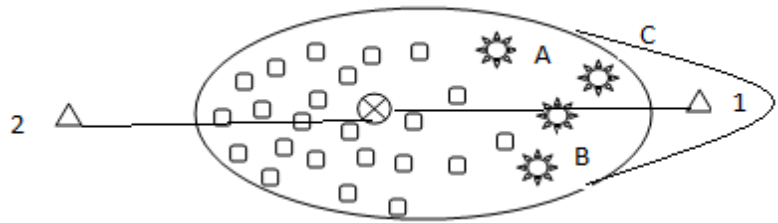

Fig.3

In the above figure 3, let point 1, be located on the sparser side and point 2 be located on the denser side of the cluster at nearly equal distances from the centroid. Point C is the farthest point in the vicinity of point 1 and Point A is the farthest point in the vicinity of point 2. Even if the point 1 is at a slightly larger distance than point 2 to the centroid it is natural to include point 1 into the cluster $C_i$ compared to point 2, due to point 1's closeness to existing members ( though farthest points) of the cluster as well as the discontinuity between point 2 to the existing boundary of the cluster on its side.

The results produced by standard partitional clustering algorithms like K-means are not in concurrence with this natural expectation as they rely upon Euclidean distance metric (ED) for discriminating data points while determining their membership into a cluster. So, the author proposes a new proximity metric called Inverse Proximity Estimate (IPE) to determine the membership of a data point into a cluster considering the data distribution within the cluster through a Bias factor in addition to normal Euclidean distance as shown below

$$IPE^{(i)}_{\Delta y} = ED\ (\overrightarrow{m_i}, \Delta y) + B$$

where, $IPE^{(i)}_{\Delta y}$ → proposed distance metric

$ED\ (\overrightarrow{m_i}, \Delta y)$ → Euclidean distance between the centroid, $\overrightarrow{m_i}$ of the cluster $C_i$ and the incoming data point, $\Delta y$.

B → Bias factor

Bias is the increment added to the conventional distance metric in view of formation of more natural clusters and better detection of outliers. It considers the unevenness / shape of the cluster reflected through a set of p- farthest points to estimate the proximity of new data points to the cluster.

### 3.3.1 Bias

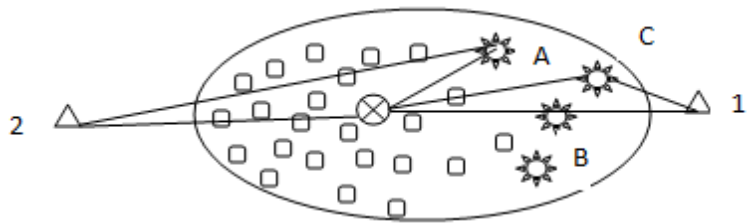

Fig.4

In Figure 4, A, B, C are the set of farthest points and it can be observed that they are located in the sparser areas of the cluster while its centroid is in the midst of dense area. Point C is the farthest point in the vicinity of point 1 and Point A is the farthest point in the vicinity of point 2. But the distance between C and point 1 is much smaller than the distance between A and point 2 which is taken as one of the factors that assess the bias and the bias increases with the distance of the new point to a farther point. So, bias is proportional to the Euclidean distance, ED between the farthest point ($q_i$) and the incoming new data point ($\Delta y$).

$$\therefore \quad B\ \alpha\ ED\ (q_i, \Delta y)$$

Another aspect to be considered is, as the distance between the centroid and the particular farthest point in the vicinity of the new point increases, elongation of the cluster with respect to that farthest point should be discouraged. This is depicted in the Figure 5 below.

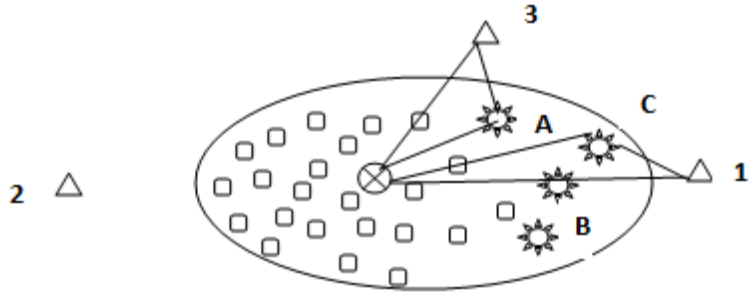

Fig.5

In the Figure 5, point 1 is in the vicinity of C and Point 3 is in the vicinity of A while both A and C are farthest points. In this context point 3 is more acceptable than point 1 into the cluster as C is farther than A to the centroid. The bias increases with the distance of a particular farthest point to its centroid. Therefore, bias is proportional to the Euclidean distance, ED between the centroid ($\vec{m_i}$) and the farthest point in the vicinity of the incoming data point ($\Delta y$).

$$\therefore \quad B \quad \alpha \quad ED\ (\vec{m_i}, q_i)$$

Hence, bias is estimated as a product of $ED\ (q_i, \Delta y)$, $ED\ (\vec{m_i}, q_i)$ mathematically from the above equations. Therefore bias is expressed as,

$$B = [\ ED\ (q_i, \Delta y)\ *\ ED\ (\vec{m_i}, q_i)\ ]$$

### 3.3.2 Proposed Proximity Metric – Inverse Proximity Estimate (IPE)

The authors have devised a new proximity estimate $IPE_{\Delta y}^{(i)}$ to determine the acceptability of an incoming data point $\Delta y$ into a possibly non-uniformly distributed cluster represented by its mean $m_i$ which is calculated as follows

$$IPE_{\Delta y}^{(i)} = ED\ (\vec{m_i}, \Delta y) + B$$

Substituting B from the above, now ($IPE_{\Delta y}^{(i)}$) is estimated as

$$IPE_{\Delta y}^{(i)} = ED\ (\vec{m_i}, \Delta y) + [\ ED\ (q_i, \Delta y)\ *\ ED\ (\vec{m_i}, q_i)]$$

For clusters with uniformly distributed points, the usual distance measures like Euclidean distance hold good for deciding the membership of a data point into a cluster. But there exist applications where clusters have non-uniform distribution of data points. In such cases, the new distance metric proposed above will be useful. The inverse proximity estimate, $IPE_{\Delta y}^{(i)}$ better recognizes the discontinuities in data space while extending the cluster boundaries as explained below.

### 3.3.3 Extendibility of the cluster boundary towards Sparse area Vs Dense area

Extendibility of a cluster to include a new point towards sparse area is based on a set of farthest points defining its boundary around the sparse area.

The points that can be included in the region nearer to the dense area by extending the boundary of the cluster towards dense area are much nearer to the centroid of this cluster only, based on the original distance metric, ED as well as the inverse proximity estimate, $IPE_{\Delta y}^{(i)}$. In other words such data points that are outside the present cluster boundary are closer to the centroid compared to existing farther members.

However, an external point which is located towards dense area at an equal distance from the centroid as the farthest points is clearly separated from the dense area with a discontinuity in between. The inverse proximity estimate, $IPE_{\Delta y}^{(i)}$ does not allow extension of the boundary of the cluster on the denser side, if there is a discontinuity between dense area and the new point, whereas such discontinuity may be overlooked by the original distance metric, ED.

Let point 5 and point 6 are equidistant from the centroid of the cluster. As shown in the Figure 6, point 5 is towards its sparse area and point 6 is towards its dense area.

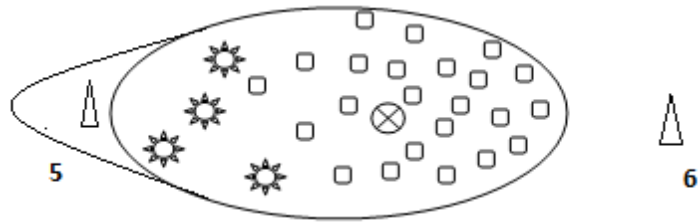

Fig. 7

Though point 6 is at an equal distance to centroid as point 5, point 6 is not acceptable into the cluster because of the discontinuity.

The only cluster members in the vicinity of point 6, with a possible discontinuity between itself and the cluster boundary, are located in the dense area. Hence, none of them are considered as farther points of the cluster. Since the sparse area containing farther points is away from point 6, Euclidean distance between the farthest point and the incoming new data point, $E_D(q_i, \Delta y)$ is high for point 6 compared to point 5, increasing the value of bias for point 6. This in turn increases the inverse proximity estimate, $IPE_{\Delta y}^{(i)}$ for point 6 compared to point 5, thereby, reducing the acceptability of point 6 into the cluster.

### 3.4. Incremental Clustering Approach with CFICA

CFICA uses the inverse proximity estimate, $IPE_{\Delta y}^{(i)}$ for effectively identifying the appropriate cluster of an incoming data point $\Delta y$. In other words, it estimates the proximity of the incoming point to a cluster based on the cluster centroid (mean $m_i$), farthest point in the vicinity of the incoming point ($q_i$) and the incoming data point ($\Delta y$).

For each cluster $c_i$, the Euclidean distance ED is calculated for the following pairs of points: centroid and incoming point $(\overrightarrow{m_i}, \Delta y)$, farthest point in the vicinity of the incoming point and incoming point $(q_i, \Delta y)$, centroid and farthest point in the vicinity of the incoming point $(\overrightarrow{m_i}, q_i)$.

Upon the arrival of a new data point $\Delta y$ to the existing database $S_D$ which is already clustered into C = {C$_1$, C$_2$, …….., C$_k$} clusters, its distance, $IPE_{\Delta y}^{(i)}$ to i$^{th}$ cluster for all i = 1 to k is calculated using the Equation specified in section 2.3.2

$$IPE_{\Delta y}^{(i)} = ED\,(\vec{m_i}, \Delta y) + [\,ED\,(q_i, \Delta y) * ED\,(\vec{m_i}, q_i)]$$

where, $ED\,(\vec{m_i}, \Delta y)$ → Euclidean distance between the points $\vec{m_i}$ and $\Delta y$

$ED\,(q_i, \Delta y)$ → Euclidean distance between the points $Q_i$ and $\Delta y$

$ED\,(\vec{m_i}, q_i)$ → Euclidean distance between the points $\vec{m_i}$ and $Q_i$

## 3.5 Finding the farthest point in the vicinity of incoming data point $\Delta y$

In order to find the farthest point $q_i$ ( data point in $Q_i$ ) which is in the vicinity of the incoming data point $\Delta y$, the Euclidean distance is calculated for that data point $\Delta y$ to each of the $p$ - farthest points of that cluster and the data point with minimum distance is designated as $q_i$ thereby, for each cluster, the point having minimum distance is taken as the farthest point $q_i$.

## 3.6 Finding the appropriate cluster

The proximity metric, IPE is used to find the appropriate cluster for the new data point, $\Delta y$. The new data point, $\Delta y$ is assigned to the closest cluster only if the calculated proximity, $IPE_{\Delta y}^{(i)}$ is less than the predefined threshold value, λ. Otherwise, the data point $\Delta y$ is not included in any of the existing clusters, but it separately forms a new singleton cluster. In such a case, the number of clusters is incremented by one.

## 3.7 Updating of Cluster Feature

From the above section, it can be seen that, whenever a new data point, $\Delta y$ is added to the existing database, that new data point may be included into any of the existing clusters or it may form a new cluster. So after the new point gets inserted, updating of $CF$ is important for further processing.

**Case 1: Inclusion of new data point into any of the existing clusters**

Whenever a new data point, $\Delta y$ is included into an already existing cluster C$_i$ its cluster feature (CF$_i$) is updated without requiring the original data points of C$_i$ and hence supports incremental update of the clustering solution.

In particular, the $n_i$, $\vec{m_i'}$, and $\vec{ss_i}$ fields of CF$_i$ are updated upon the arrival of a new data point $\Delta y$ into the cluster C$_i$ as given below:

(1) $n_i = n_i + 1$

(2) $\vec{m_i}' = \dfrac{n_i * \vec{m_i}' + \Delta y}{n_i}$

(3) $\vec{ss_i} = \vec{ss_i}$ + squared components of $\Delta y$.

However, the $Q_i$ representing the p-farthest points and the centroid of the previous snapshot $\vec{m_i}'$, were kept without any changes until the next periodical refresh.

**Case 2: New data point forms a singleton cluster**

If the new data point forms a new cluster separately, the cluster feature ($CF_i$) has to be computed for the new cluster containing the data point $\Delta y$. $CF_i$ for the new cluster contains the following information:

(1) Number of data points. Here n=1, as it is a singleton cluster,
(2) Mean of the new cluster; In this case, $\vec{m_i} = \vec{m_i}' = \Delta y$
(3) The list of p-farthest points includes $\Delta y$ only.
(4) Squared sum is squared sum of components of $\Delta y$.

Finding the appropriate cluster to incorporate the new data point $\Delta y$ and updating of the cluster feature after adding it to the existing cluster or forming a separate cluster are iteratively performed for all the data points in the incremental database $\Delta S_D$.

### 3.8 Merging of closest cluster pair

Once the incremental database $\Delta S_D$ is processed with CFICA, the need to merge the closest clusters may arise. A merging strategy is used to maintain reasonable number of clusters with high quality. Therefore, merging is performed when the number of clusters increases beyond 'k' ( k in k-means) while ensuring that increase in variance which indicates error is minimum due to merging. It is intuitive to expect an increase in the error with the decrease in the number of clusters. A closest cluster pair is considered for merging if only the Euclidean distance between the centroids of the pair of clusters is smaller than user defined merging threshold (θ).

The procedure used for the merging process is described below:

Step 1: Calculate the Euclidean distance ED between every pair of cluster centroids ($\vec{m_i}$).

Step 2: For every cluster pair, with Euclidean distance, ED less than the merging threshold value, $\theta$ ( ED $\ll \theta$ ), find increase in variance ($\sigma^2$) as described in section 3.2.

Step 3: Identify the cluster pair ($C_i$, $C_j$) with minimum increase in variance.

International Journal of Data Mining & Knowledge Management Process (IJDKP) Vol.3, No.5, September 2013

Step 4 : Merge $C_i$ and $C_j$ to form the new cluster, $C_k$ and compute the Cluster Feature  for $C_k$ and delete $C_i$ and $C_j$ along with their *CF's*.

Step 5 : Repeat steps 1 to 4 until no cluster pair is mergable or until the value of 'k' is adjusted.

**Computing Cluster Feature of merged cluster**

After merging the closest cluster pair, now the Cluster Feature ( CF ) has to be computed for the merged cluster. The CF of the merged cluster is calculated by:

    (1)    Adding the number of data points in both the clusters
    (2)    Calculating the mean of the merged cluster
    (3)    Finding the p-farthest points of the merged cluster
    (4)    Incrementing the squared sum.

In particular, when the $i^{th}$ cluster with $CF_i = \{ n_i, \vec{m_i}, \vec{m_i'}, Q_i, \vec{ss_i} \}$ and $j^{th}$ cluster with $CF_j = \{ n_j, \vec{m_j}, \vec{m_j'}, Q_j, \vec{ss_j} \}$ are merged to form $k^{th}$ cluster, its cluster feature $CF_k$ is calculated as given below:

1) $n_k = ( n_i + n_j )$

2) $\vec{m_k} = \dfrac{(n_i * \vec{m_i}) + (n_j * \vec{m_j})}{n_k}$

3) $\vec{m_k'} = \vec{m_k}$

4) p-farthest points are selected from farthest points in $q_i$ and $q_j$.

5) $\vec{ss_k} = \vec{ss_i} + \vec{ss_j}$

Hence the Cluster Feature of the new cluster 'k' which is formed by merging two existing clusters 'i' and 'j' is determined as a function of $CF_i$ and $CF_j$.

$$CF_k = f ( CF_i , CF_j)$$

Thus Cluster Feature provides the essence of the clustered data points thereby avoiding explicit referencing of individual objects of the clusters which may be maintained in the external memory space.

### 3.9 Need for Cluster Refresh

The addition of new data points into some of the existing clusters naturally results in change of mean. So, the set of p-farthest points of cluster $C_i$, need not be the p-farthest points of cluster $C_i'$, modified version of $C_i$. Whether or not, the p-farthest points of the cluster are the same, depend upon the recently added data points. Ideally, a new set of p-farthest points have to be computed for the incremented cluster $C_i'$. Calculating p-farthest points again, every time the cluster gets updated, is not an easy task as it involves recalculation of the Euclidean distance, ED for every data point in the incremented cluster $C_i'$ with the updated mean, $\vec{m_i'}$.

101101

For pragmatic reasons, it was suggested to refresh the $CF_i$, only in case it deviates significantly from its original value indicating concept drift [Chen H.L et al. 2009]. After processing a new chunk of data points from $\Delta S_D$ the deviation in mean is calculated as follows

$$\text{Deviation in mean} = \frac{\vec{m_i'} - \vec{m_i}}{\vec{m_i}}$$

Those clusters with deviation in mean greater than δ, based on a threshold are identified as clusters with significant concept – drift and hence needs to be refreshed. The process of cluster refresh involves finding a new set of p-farthest points for such clusters. For the remaining clusters, the same set of p-farthest points along with the old mean value, $m_i$ is maintained. It may be noted that $m_i$ in the $CF_i$ always represents the centroid based on which the p-farthest points are identified. Hence needs to be changed whenever new set of p-farthest points are identified. The above process of cluster refresh is applicable only to the incremented cluster $C_i'$ due to inclusion of additional data points.

## 4. EXPERIMENTAL ANALYSIS

CFICA has been implemented using the Iris [13], wine [14] and yeast [15] datasets from the UCI machine learning repository. All the datasets were preprocessed. Iris and Wine datasets do not have any missing values but Yeast dataset has some missing attribute values and such instances were ignored. After preprocessing, yeast dataset had 1419 instances. Normally datasets used for the purpose of analysis may contain too many attributes, which may or may not be relevant. Therefore dimensionality reduction has been done on Wine and Yeast datasets as they contain a considerable number of attributes. So the number of attributes came down from 13 to 3 for wine, from 8 to 5 attributes for yeast datasets.

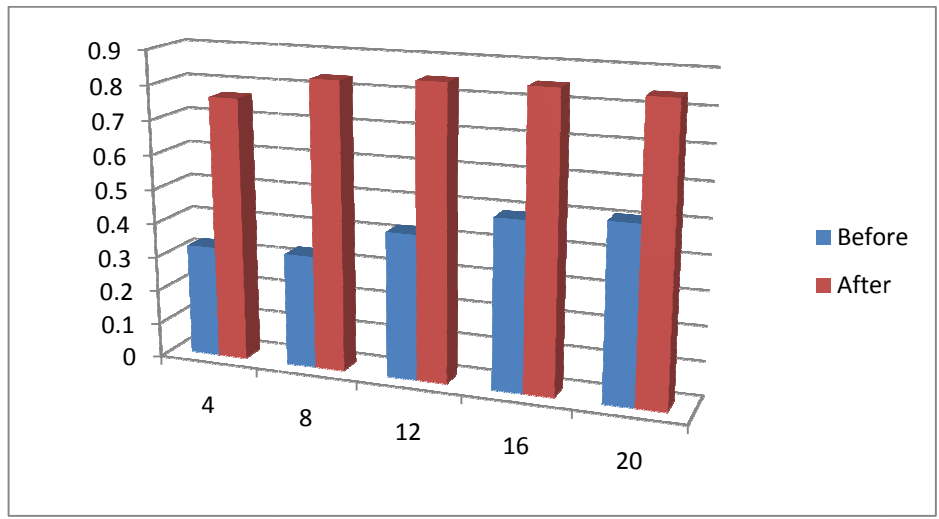

Fig 1. Purity of Wine dataset before and after dimensionality reduction

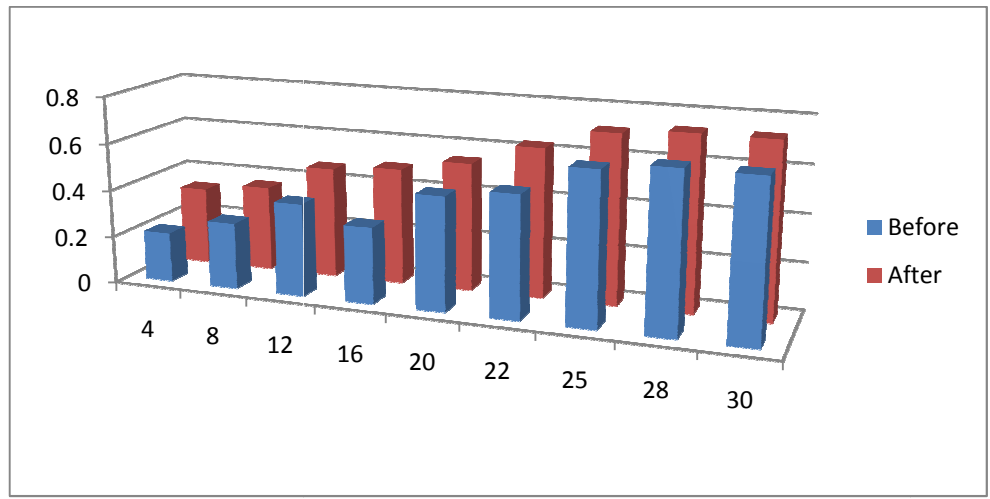

Fig 2. Purity of Yeast dataset before and after dimensionality reduction

It can be seen from the above figures that the cluster purity increases after dimensionality reduction has been done.

### 4.1 Cluster formation

The above datasets are processed dynamically by dividing the data points in each dataset into chunks. For example, the Iris dataset is divided into 4 chunks (Iris1 consisting of 75 instances, Iris2, Iris3 and Iris4 consisting of 25 instances each). Initially, the first chunk of data points is given as input to the k-means algorithm for initial clustering i.e Iris1. It generates k number of clusters. Then, the cluster feature is computed for those k initial clusters. The next chunk of data points is input to the new approach incrementally. For each data point from the second chunk, the Inverse Proximity Estimate (IPE) is computed and the data points are assigned to the corresponding cluster if the calculated distance measure is less than the predefined threshold value, $\lambda$. Otherwise, it forms as a separate cluster. Subsequently, the cluster feature is updated for each data point. Once the whole chunk of data points is processed, the merging process is done if only the Euclidean distance between the centroids of the pair of clusters is smaller than user defined merging threshold, $\theta$. Here, $\lambda = 10$ and $\theta = 4$. Finally, the set of resultant clusters are obtained from the merging process and the cluster solution is updated incrementally upon receiving the later chunks of data Iris3 and Iris4. Similarly, wine and yeast datasets are also divided into 4 chunks as follows: wine1 - 100 instances, wine2 and wine3 - 25 instances each, wine4 - 28 instances, yeast1 - 700 instances, yeast 2 – 350 instances, yeast3 - 200 instances, yeast4 - 168 instances and handled as above.

### 4.2 Metrics in which performance is estimated

Validation of clustering results is important. Therefore, the Purity Measure has been used to evaluate the clustering results obtained. A cluster is called a pure cluster if all the objects belong to a single class. The purity measure described in [16] [17] has been used for evaluating the performance of CFICA. The evaluation metric used in CFICA is given below,

$$\text{Purity} = \frac{1}{N} \sum_{i=1}^{T} X_i$$

where, $N \rightarrow$ Number of data points in the dataset

$T \rightarrow$ Number of resultant cluster

$X_i \rightarrow$ Number of data points of majority class in cluster $i$

The purity of the resultant clusters is calculated by changing the k-value (order of initial clustering) and is graphically presented in Figures 3 to 5.

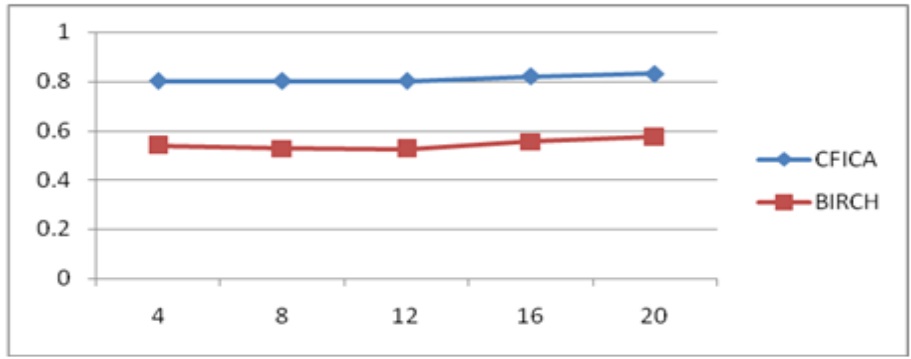

Fig 3. Purity vs. number of clusters ($k$) for Iris dataset.

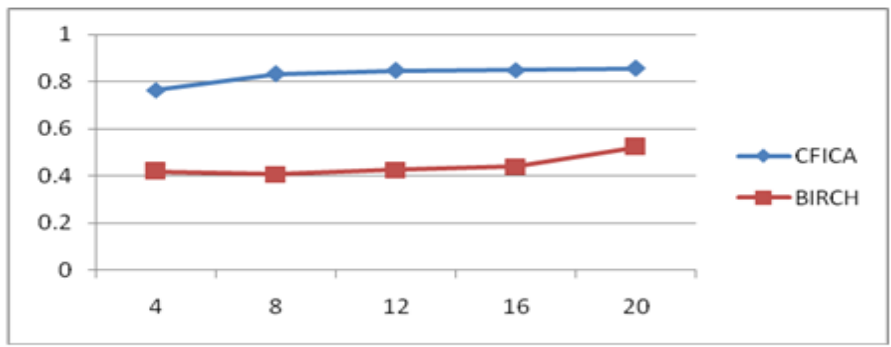

Fig 4. Purity vs. number of clusters ($k$) for Wine dataset.

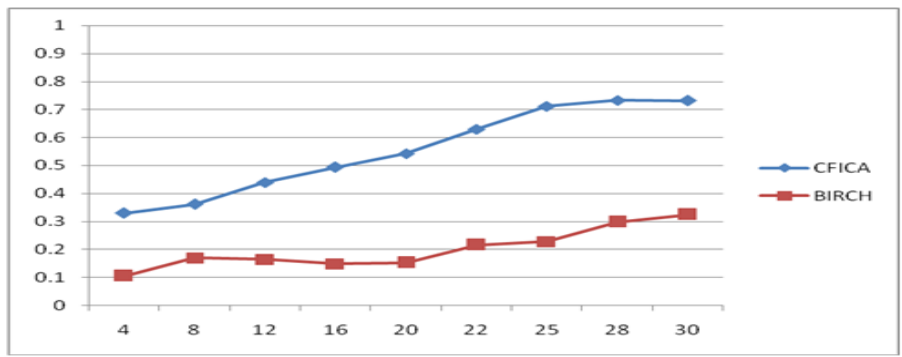

Fig 5. Purity vs. number of clusters ($k$) for Yeast dataset.

The above results demonstrate that CFICA performs better than BIRCH in terms of purity. BIRCH algorithm is unable to deliver satisfactory clustering quality if the clusters are not spherical in shape because it employs the notion of radius or diameter to control the boundary of a cluster.

## 5. THE PSEUDO-CODE FOR CFICA

The various steps constituting the proposed algorithm CFICA is listed out in the form of pseudo-code. CFICA accepts the clustering solution for existing database $S_D$ in the form of cluster features as input and incrementally updates the set of cluster features in $\Delta S_D$ in accordance with newly arrived chunk of data points.

**Input**

$S_D \rightarrow$ Initial set of CF's for $S_D$.

$\Delta S_D \rightarrow$ Set of data points added to $S_D$

K $\rightarrow$ Number of clusters

θ $\rightarrow$ Predefined merging threshold value

δ $\rightarrow$ Allowed deviation in mean

λ $\rightarrow$ User defined radius threshold

**Output**

Set of CF's { $CF_1, CF_2, \ldots, CF_K$ } for ( $S_D + \Delta S_D$)

**Variables**

$\Delta y \rightarrow$ New data point

$CF_i \rightarrow$ Cluster Feature of i[th] cluster

$IPE_{\Delta y}^{(i)} \rightarrow$ Inverse Proximity Estimate between $\Delta y$ and i[th] cluster

ED $\rightarrow$ Euclidean distance

$n_i \rightarrow$ Number of data points

$\vec{m_i} \rightarrow$ Mean vector of the cluster

$\vec{m_i'} \rightarrow$ New mean vector of the cluster

$Q_i \rightarrow$ List of p-farthest points of i[th] cluster

$q_i \rightarrow$ The data point in $Q_i$ which is closest to $\Delta y$

$\vec{ss_i} \rightarrow$ Squared sum vector

**Algorithm**

1) Compute Cluster Feature for every cluster $C_i$

$$CF_i = \{ n_i, \overrightarrow{m_i}, \overrightarrow{m_i}, Q_i, \overrightarrow{ss_i}\}$$

a = k;

2) For each data point $\Delta y \in \Delta S_D$

a) For i from 1 to a

Calculate Inverse Proximity Estimate of $\Delta y$ to the i$^{th}$ cluster

$$IPE_{\Delta y}^{(i)} = ED\,(\overrightarrow{m_i}, \Delta y) + [\,ED\,(q_i, \Delta y) * ED\,(\overrightarrow{m_i}, q_i)]$$

b) Find suitable cluster j = arg min$_i$ { $IPE_{\Delta y}^{(i)}$ }

if ( $IPE_{\Delta y}^{(i)} < \lambda$ )

{

insert $\Delta y$ into its closest cluster j and update its CF as follows:

$$CF_j = \{\,(n_j+1), \overrightarrow{m_j}, \frac{n_j * \overrightarrow{m_j} + \Delta y}{n_j + 1}, q_j, (\overrightarrow{SS_i} + squared\ components\ of\ \Delta y)\}$$

if ( deviation in mean = $\frac{|\overrightarrow{m_j'} - \overrightarrow{m_j}|}{|\overrightarrow{m_j}|} > \delta$ )

Read the data points of j$^{th}$ cluster to recompute $CF_j$.

}

else create a new cluster

{

increment a

insert $\Delta y$ into a$^{th}$ cluster

$$CF_a = \{1, \overrightarrow{m_a}, \overrightarrow{m_a}, \Delta y, squared\ components\ of\ \Delta y\}$$

}

International Journal of Data Mining & Knowledge Management Process (IJDKP) Vol.3, No.5, September 2013

    3)       If ( a > k ) merge cluster pair

        (i) Compute ED between every pair of cluster centroids based on $\vec{m_i}$.

        (ii) For cluster pairs with ( ED $\ll \theta$ ), find increase in variance.

        (iii) Merge the cluster pair with minimum increase in variance

        Decrement a

        (iv) Recalculate CF of merged cluster

$$CF_k = \{ (n_i + n_j), \frac{(n_i * \vec{m_i}) + (n_j * \vec{m_j})}{n_k}, \vec{m_k} = \vec{m'_k}, q_{k,}, \vec{SS_k} = \vec{SS_i} + \vec{SS_j} \}$$

        Find its ED to other clusters.

        (v) Repeat steps (i) to (iv) until ( a = k )

    4)       Print cluster solution as the set of cluster features.

        Wait for the arrival of new chunk of data points upon which call CFICA again.

## 6. CONCLUSIONS

An incremental clustering algorithm called Cluster Feature-Based Incremental Clustering Approach for numerical data (CFICA) which makes use of Inverse Proximity Estimate to handle entities described in terms of only numerical attributes was developed and evaluated. Cluster Feature while being compact includes all essential information required for maintenance and expansion of clusters. Thus CFICA avoids redundant processing which is the essential feature of an incremental algorithm. The performance of this algorithm, in terms of purity is compared with the state of art incremental clustering algorithm namely BIRCH on different bench mark datasets.

International Journal of Data Mining & Knowledge Management Process (IJDKP) Vol.3, No.5, September 2013

## AUTHORS

**A.M.Sowjanya** received her M.Tech. Degree in Computer Science and technology from Andhra University. She is presently working as an Assistant Professor in the department of Computer Science and Systems Engineering, College of Engineering (Autonomous), Andhra University, Visakhapatnam, Andhra Pradesh, India. She submitted her Ph.D in Andhra University. Her areas of interest include Data Mining, incremental clustering and Database management systems.

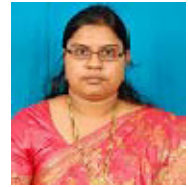

**M.Shashi** received her B.E. Degree in Electrical and Electronics and M.E.Degree in Computer Engineering with distinction from Andhra University. She received Ph.D in 1994 from Andhra University and got the best Ph.D thesis award. She is working as a professor of Computer Science and Systems Engineering at Andhra University, Andhra Pradesh, India. She received AICTE career award as young teacher in 1996. She is a co-author of the Indian Edition of text book on "Data Structures and Program Design in C" from Pearson Education Ltd. She published technical papers in National and International Journals. Her research interests include Data Mining, Artificial intelligence, Pattern Recognition and Machine Learning. She is a life member of ISTE, CSI and a fellow member of Institute of Engineers (India).

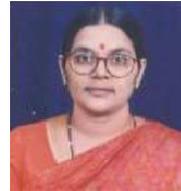